\documentclass[twocolumn,secnumarabic,amssymb, nobibnotes, aps, prd]{revtex4-2}
\usepackage{graphicx}

\begin{document}

\title{Predicting Dynamics of Transmon Qubit-Cavity Systems with Recurrent Neural Networks } 

\begin{abstract}
Developing accurate and computationally inexpensive models for the dynamics of open-quantum systems is critical in designing new qubit platforms by first understanding their mechanisms of decoherence and dephasing. Current models based on solutions to master equations are not sufficient in capturing the non-Markovian dynamics at play and suffer from large computational costs. Here, we present a method of overcoming this by using a recurrent neural network to obtain effective solutions to the Lindblad master equation for a coupled transmon qubit-cavity system. We present the training and testing performance of the model trained a simulated dataset and demonstrate its ability to map microscopic dissipative mechanisms to quantum observables.    
\end{abstract}

\author{Nima Leclerc}%
\email{nleclerc@seas.upenn.edu}
\affiliation{Department of Electrical and Systems Engineering, University of Pennsylvania, Philadelphia, Pennsylvania 19104, USA}
\date{May 2021}%
\maketitle

\section{Introduction}
The development of both deterministic and stochastic models for open quantum systems have been powerful in revealing the physical mechanisms leading to dissipation and decoherence in materials and devices for quantum information processing \cite{rotter}. In particular, quantum-dynamic models that account for non-unitary evolution and are driven by simulated and experimental data will accelerate the engineering of new qubit designs by establishing a relationship between the underlying dissipative processes and device performance.  Unfortunately, conventional models including the Lindblad master equation \cite{manzano} and correlated cluster expansions \cite{yang} are limited in their ability to model complex open quantum systems to high accuracy efficiently due to the exponential computational cost of solving for quantum trajectories and the inherent Markovian assumptions. For instance, the Lindblad master equation alone is unable to accurately model the $T_{2}$ decoherence times in diamond nitrogen vacancy center qubits, as the model does not account for the history of evolution of the nuclear-spin bath causing dissipation \cite{viktor}.  
\par  Relying only on first-principles  models (e.g., using only the Lindblad master equation) for dissipative quantum dynamics will not be sufficient in furthering designs of new qubits for the model assumptions mentioned, high computational cost, and lack of generalization. This motivates the need for a data-driven approach.
\par Since the early 2010s, recurrent neural networks (RNNs) have become a powerful tool for recognition and prediction of patterns in audio signals, which has been responsible for the proliferation of speech-recognition systems over the past decade \cite{lipton}. RNNs work particularly well with time-series data (e.g.,  audio signals) as the networks are parameterized by both time-dependent and time dependent weights, the networks are trained on multiple instances of time series, and are inherently non-Markovian as the prediction of $x_t$ at time $t$ is determined from a probability distribution $\mathcal{P} (x_t | x_{t-1}, \cdots, x_{t-N})$ depending on the entire history of $N$ previous time steps.  
\par Several groups have  successfully used RNNs to predict quantum trajectories capturing non-unitary evolution using both simulated and experimental datasets \cite{banchi,flurin}. These works have demonstrated the models' abilities to capture the input-output relation between system parameters and trajectories without a closed analytical form. 
Banchi \emph{et al.}  used the RNN model to effectively solve the Lindblad master equation for toy systems under different quantum state initializations, while Flurin \emph{et al.} used the model to predict quantum trajectories for transmon qubits under different microwave control sequences. However, there are no efforts in the literature that investigate the relationship between microscopic dissipative processes and measured quantum trajectories which would otherwise be challenging to solve from first-principles in a generalized way. In this work, we use a transmon qubit coupled to a superconducting cavity as a model system to investigate the relationship between dissipative processes of the system and the measured quantum dynamics using an RNN model.  The objective of the work is to generate an input-output relationship of dissipative mechanisms to time traces of measurement outcomes (e.g. $\langle \mathbf{\sigma} \rangle, \langle \hat{n} \rangle$). This paper is organized as follows: Section II will overview the physical qubit-cavity system that we consider, Section III will introduce the master equation that we solve to generate our training set, Section IV will discuss the architecture of our RNN, Section V will discuss our specific implementation of generating data and training, Section VI will overview our results, and we summarize our outcomes and path forward in Section VII.    
\section{Transmon Qubit Dynamics}
Often, it is instructive to study the quantum dynamics and dissipation of a qubit by investigating its interactions with a cavity. In the case of our model system, we consider superconducting transmon qubit (Cooper pair box) coupled to a 1D transmission line resonator which plays the role of a microwave cavity.  As derived by Blais \emph{et al.},  we can model  the composite system using the modified Jaynes-Cummings Hamiltonian $\mathcal{H}_{sys}$  in Eqt.(1)  \cite{blais}.  
\begin{equation} 
\mathcal{H}_{sys} =  \hbar((\Delta_c a^{\dagger} a  - \frac{1}{2} \Delta_q \sigma_z   + g(a^{\dagger} \sigma^{-}  +  a \sigma^{+} )  +  \eta (t) (a^{\dagger} + a ))  
\end{equation} 
Where $\Delta_c = \omega_c - \omega_d$ is the detuning of the cavity with $\omega_c$ as the cavity resonant frequency and $\omega_d$ as the microwave drive frequency, $\Delta_q =\omega_q - \omega_d$ is the detuning of the transmon qubit with $\omega_q$ as the frequency of the qubit ($ \hbar \omega_q = E_1 - E_0 $ , with $E_0$ and $E_1$ being the ground state and first excited state energies of the qubit), $g$ is the vacuum Rabi frequency of the system, $\eta (t)$ is the amplitude of the drive, $a$($a^{\dagger}$) is the lowering (raising) operator for microwave photon number in the cavity, $\sigma^{-}$( $\sigma^{+}$) is the lowering (raising) operator of the qubit from the excited to ground (ground to excited state), and $\sigma_z$ is the Pauli-$z$ operator in the excited-ground state basis. For this study, we take $\omega_r$, $\omega_c$, and $g$ to be fixed parameter of the fabricated devices and set $\eta(t)=\omega_q=0$. We will discuss the relevant dissipative processes and their associated parameters in the following section.    
\section{Quantum Trajectories}
Having defined the Hamiltonian $\mathcal{H}_{sys}$ in Eqt.(1) describing the coupled system, we can move forward in  modeling the system's dissipative dynamics using a Linbladian formalism. This is necessary in generating the dataset for  the RNN model, as points in our training set will distinguished by unique dissipation parameters found in the master equation. In particular, consider the master equation in Eqt.(2).   
\begin{equation} 
\dot{\rho}= -\frac{i}{\hbar} [\mathcal{H}_{sys}, \rho] + \kappa \mathcal{D} [a] \rho + \Gamma_1 \mathcal{D} [\sigma^{-} ]  \rho + \frac{\Gamma_{\phi} }{2}  \mathcal{D} [\sigma^z ] \rho   
\end{equation}   
Eqt. (2) models the non-unitary evolution of our coupled system due photon loss of the cavity and dephasing of the qubit, where the solutions $\rho$ correspond to time traces of the density matrix over the decay. In Eqt. (2),  $\rho$ is the state's density matrix in the cavity number-qubit state basis ($\rho = \rho_{c,n} \otimes  \rho_{q,i} $),  $\kappa$ is the rate of photon leakage in the cavity, $\Gamma_1$ is the qubit decay rate (from excited to ground state), and $\Gamma_{\phi}$ is the dephasing rate of the qubit.  For each of the these non-unitary processes, we have a Lindbladian superoperator of the form $\mathcal{D} [A] \rho  = \frac{1}{2} (A \rho A^{\dagger} - \{A^{\dagger} A, \rho \}  )  $ for some operator $A$. In generating our training set, we will allow the parameters $\kappa$ and $\Gamma_{\phi}$ to vary, keeping the rest fixed. Varying these parameters at the input, we collect time traces of the outcomes $\langle \sigma_z \rangle $, and $\langle a^{\dagger} a \rangle $. 

\section{Recurrent Neural Network}  
RNNs are class of neural networks which comprise recurrent layers, allowing for the time state of an input to be maintained as it propagates through a network $f$. Labeling each layer of the network with $l$,  the output of a layer takes the form $s_t^{l+1} = f(W^l  s_t^l + w^l , s^{l+1}_{t-1} )$ where $s_t^l$ represents the input at time $t$ for layer $l$ and  $W^l$ and  $w^l$ are time-independent trainable weights. It's noteworthy that the state at a given time $t$ depends on all of the previous time states.  In this work, the input states $\{ s_t \} $ represent the input dissipation parameters $\{ \kappa_t, \Gamma_{\phi, t} \} =\{ \kappa_{t=0} , \Gamma_{\phi, t=0} \}$ while the network outputs correspond to the observable  $\langle \sigma_z \rangle $, and $\langle a^{\dagger} a \rangle$. The job of the RNN model is to solve an effective master equation given by $\dot{\rho}   = \mathcal{L}_{\leq t }^{RNN}  [\rho(t)   ]   $ where the operator $\mathcal{L}_{\leq t, W }^{RNN}$ is a superoperator parameterized by weights $W$ that acts on the density matrix $\rho$ of the system. The process of training the network solves for the weights $W$, while evaluating the the network during forward-pass produces a time trace $\rho (t)$ corresponding the network's predictions. 
  \par In this work, we adopt the architecture adopted by Banchi \emph{et al.} \cite{banchi} and given in Figure 1. The network takes in a set of parameters $\{W_{d,t} \}_N  \in $ such that $W_{d,t} \in  \mathcal{R}^{ d \times N_t} $  with $d$ as the type of input parameter, $N$ as the number of settings, and $N_t$ as the length of the time sequence for each point.     
\begin{figure}[ht]
\center{\includegraphics[width=0.4\textwidth]
     {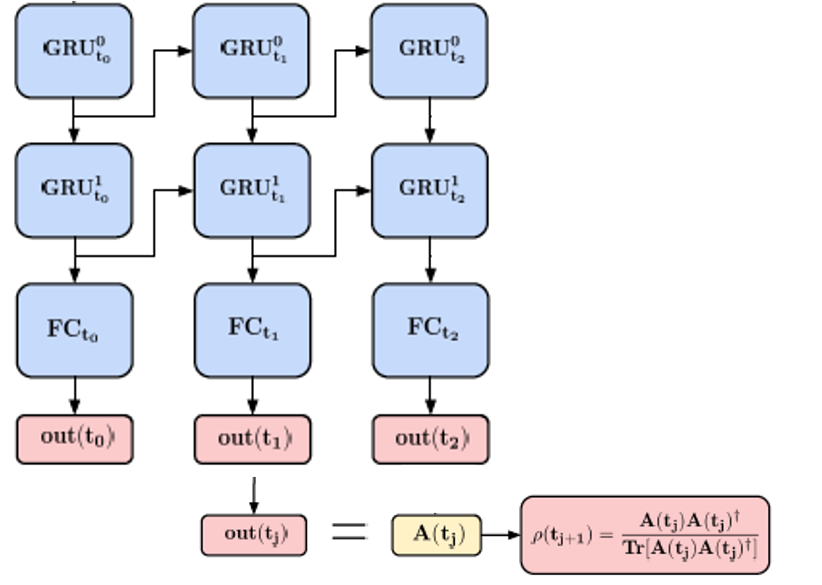}}
   \caption{\label{fig1: network}  RNN implementation for prediction of trajectories $\rho (t_j)$ adopted from Banchi \emph{et al.} \cite{banchi}.  }
 \end{figure}
The network comprises two gated recurrent unit layers (GRUs) [see references \cite{wu} for details on GRUs], followed by a linear fully connected layer, and terminated with a ReLU activation function. However, the outputs of the network are a high-dimensional representation of the desired observable given by $A (t_j)$. This high-dimensional representation can be mapped to the true observable $\rho (t_j)$ from Eqt. (3).  
\begin{equation} 
\rho (t_{j+1})  = \frac{ A( t_j) A(t_j)^{\dagger}  }{ \text{Tr}  [A( t_j) A(t_j)^{\dagger}   ]  }  
\end{equation} 
Provided that each parameter setting and time-step the network produces an output $\hat{\rho}_{\alpha} (t_j)$ and has a corresponding output in the training set  $\rho_{\alpha} (t_j)$, we can write the the cost function of the model as in  Eqt. (4). 

\begin{equation} 
L  = \frac{1}{ N_b N_T}  \sum_{\alpha = 1}^{N_b }   \sum_{j = 1}^{N_T }  || \hat{\rho}_{\alpha} (t_j) -  \rho_{\alpha} (t_j) ||_2^2  
\end{equation}   
Given that $N_b$ is the size of a mini-batch and $N_T$ is the number of time-steps, we can minimize $L$ using a variant  of stochastic gradient descent to find the optimal parameter set $\{W_{d,t} \}_N$ obtained at the end of training. In the following section, we will discuss our specific implementation.    
\section{Methods and Model Description}
We first generated our training and testing datasets by solving the Eqt.(2) numerically for the coupled transmon-cavity system, where we fix the Hamiltonian parameters $\omega_c = 45.12$ GHz, $\omega_q =27.07$ GHz , $\Gamma_1  = 27$ MHz,  and $g = 690.8$ MHz  in Eqt.(1) to experimental parameters of a device found in the literature \cite{blais}. Working in a Fock basis of size 5 and initializing the qubit in its first excited state, we perform 1,000 simulations over a grid of parameters $\kappa$ and  $\Gamma_{\phi}$  given by $ \{\kappa \} \times  \{ \Gamma_{\phi} \} = \{  [ 75, 130] \text{  MHz} \} \times  \{  [25, 100] \text{  MHz} \}  $ under no external drive.  We then obtain a set of 1,000 time traces of $\langle \sigma_z \rangle $ and $\langle a^{\dagger} a \rangle $, each with $N_T = 100$ time steps.  From here, we performed a standard 80\%/20\%  training/testing split of the data set producing data sets $\mathcal{D}_{train} \in \mathcal{R}^{N = 800 \times N_{T} = 100 \times d = 3 }$ and $\mathcal{D}_{test} \in \mathcal{R}^{N = 200 \times N_{T} = 100 \times d = 3 }$.   
\par Provided our training and test data sets, we proceeded to train the network. Using the architecture outlined in Section IV, we trained our network on $\mathcal{D}_{train}$ using layers with hidden dimension of 40, using the Adam optimizer and setting the  learning rate to 0.0001 and mini-batch size to $N_b = 50$.  We then trained over 100 epochs until convergence was reached for each observable. In the validation stage, the model predictions $\hat{\rho_{\alpha}}$ were compared against the true outcomes $\rho_{\alpha}$ in $\mathcal{D}_{test}$ and test errors were extracted. 

\section{Results and Discussion}
\par Following the methods discussed in the previous section, we first simulated our dataset and then split into the training and test sets. As an example, we plot time traces from the training set for the measurement outcomes  $\langle \sigma_z \rangle$ and $\langle a^{\dagger} a \rangle$ with the input parameters  $\kappa = 100$ MHz and  $\Gamma_{\phi} = 50$ MHz in Figure 2. Clearly, these results indicate that dissipative processes are present for both the cavity and the qubit as the photon number in orange drops off from its max value of 5 to 3 over the dissipation time.  Likewise, the qubit quickly decays from its excited to its ground state as indicated by the curve in blue. However, the parameters $\kappa$ and $\Gamma_{\phi}$ control the size of the initial ripples and the steepness of the decay.

\begin{figure}[ht]
\center{\includegraphics[width=0.3\textwidth]
     {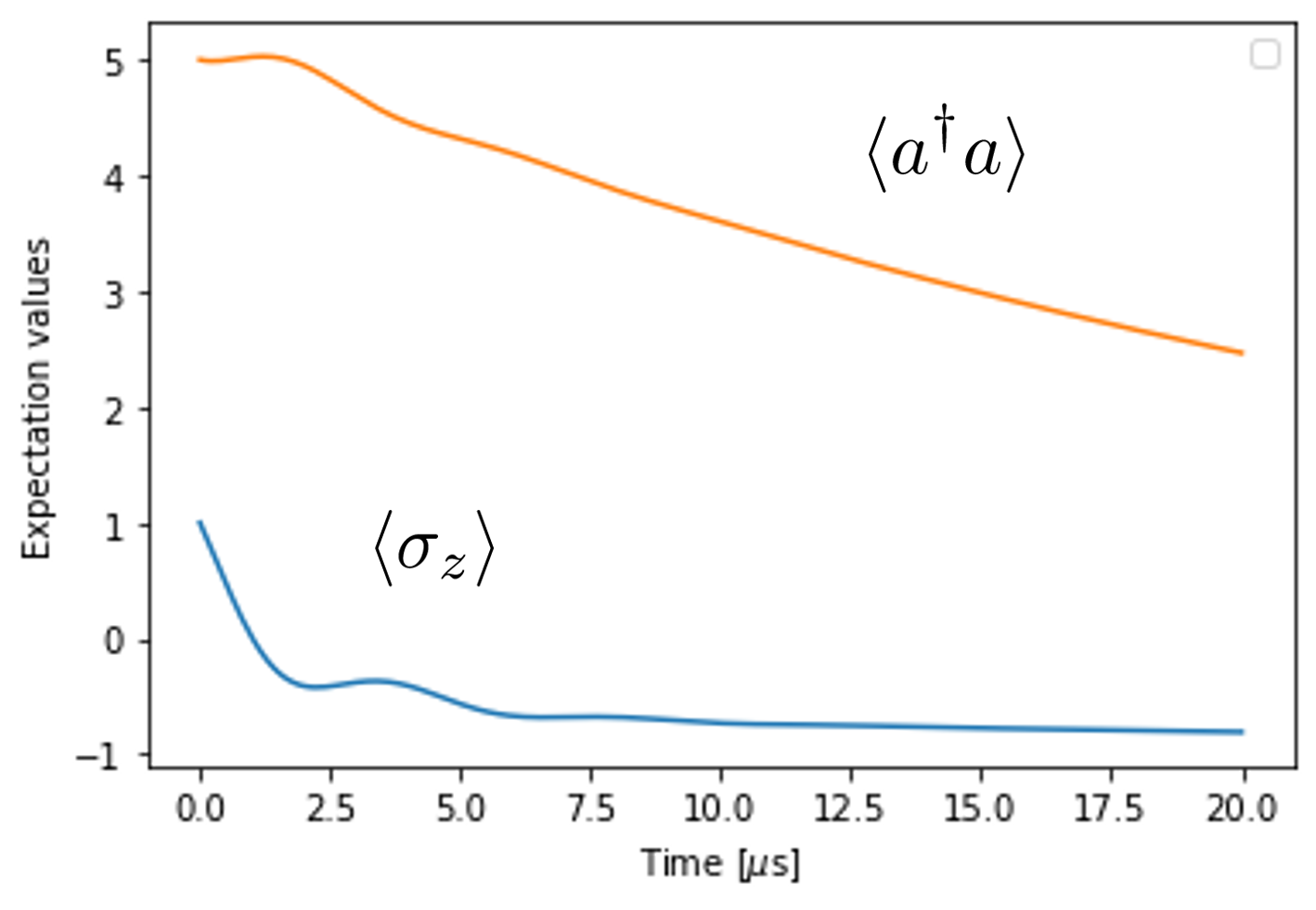}}
   \caption{\label{fig2: training set}  Example of time traces of $\langle \sigma_z \rangle$ and $\langle a^{\dagger} a \rangle$   from training set for $\kappa = 100$ MHz and  $\Gamma_{\phi} = 50$ MHz.}
 \end{figure}
 \par The network's training loss averaged over the observables  $\langle \sigma_z \rangle$ and $\langle a^{\dagger} a \rangle$  (Eqt. 4)  is plotted in Figure 3. From this, we observe that convergence in the training loss/error is obtained only after 20 epochs. However, we extracted the model parameters at 50 epochs.  Overall, this convergence demonstrates very good training performance and the training accuracy we obtained after the 50th epoch was 96.1 \%. Going further, we performed validation on our test set and averaged to get a net validation error of 94.3 \%. Although these numbers indicate high test and training accuracy, a stronger effort needs to be dedicated to evaluating the bias and variance of the model to ensure better generalization performance.

 \begin{figure}[ht]
\center{\includegraphics[width=0.3\textwidth]
     {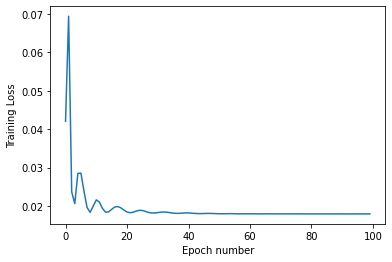}}
   \caption{\label{fig2: training performance} Average loss function for  $\langle \sigma_z \rangle$  and $\langle a^{\dagger} a \rangle$ over course of training for 100 epochs. }
 \end{figure}

\section{Conclusion and Outlook}
In this work, we presented the first demonstration of a recurrent neural network (RNN)  used to predict time traces of quantum mechanical observables from  dissipative microscopic mechanisms. We considered a superconducting transmon-cavity device as a model system to study the interplay between the time dependent observables  $\langle \sigma_z \rangle$,  and $\langle a^{\dagger} a \rangle$ and dissipation parameters $\kappa$ and $\Gamma_{\phi}$. We trained our RNN to make these predictions with sufficient training and test accuracy. Although our model was only implemented on a single model system , this model can be extended to other more complex open quantum systems including arrays of entangled qubits coupled to one another and be trained on experimental time-traces which are non-Markovian in nature. Going beyond using our proposed method to obtain non-Markovian solutions to the Lindblad master equation, this model will be an effective tool in identifying the dissipative mechanisms that are most detrimental to experimental devices, allowing one to work between the model, qubit device fabrication, and measurement to improve overall performance.

\end{document}